\documentclass[]{pasj01}
\draft

\begin{document}
\Received{2017/5/31}
\Accepted{2017/10/02}

\title{The effect of photoionising feedback on star formation in isolated and colliding clouds}
\author{Kazuhiro \textsc{Shima}\altaffilmark{1,}$^{*}$}
\author{Elizabeth J. \textsc{Tasker}\altaffilmark{2}}
\author{Christoph \textsc{Federrath}\altaffilmark{3}}
\author{Asao \textsc{Habe}\altaffilmark{1}}
\altaffiltext{1}{Department of Physics, Faculty of Science, Hokkaido University, Kita 10 Nishi 8 Kita-ku, Sapporo 060-0810, Japan}
\altaffiltext{2}{Institute of Space and Astronomical Science, Japan Aerospace Exploration Agency, Yoshinodai 3-1-1, Sagamihara, Kanagawa, 252-5210, Japan}
\altaffiltext{3}{Research School of Astronomy and Astrophysics, Australian National University, Canberra, ACT 2611, Australia}
\email{shima@astro1.sci.hokudai.ac.jp}


\maketitle

\begin{abstract}
  We investigate star formation occurring in idealised giant molecular clouds, comparing structures that evolve in isolation versus those undergoing a collision. Two different collision speeds are investigated and the impact of photoionising radiation from the stars is determined. We find that a colliding system leads to more massive star formation both with and without the addition of feedback, raising overall star formation efficiencies (SFE) by a factor of 10 and steepening the high-mass end of the stellar mass function. This rise in SFE is due to increased turbulent compression during the cloud collision. While feedback can both promote and hinder star formation in the isolated system, it increases the SFE by approximately 1.5 times in the colliding case when the thermal speed of the resulting HII regions matches the shock propagation speed in the collision.
\end{abstract}

\section{Introduction}

While star formation is known to occur in the giant molecular clouds (GMCs) that form the coldest part of the interstellar medium (ISM), it remains far from clear what factors control the conversion from cloud gas into stars. Simple considerations tell us that the GMCs cannot be undergoing free-fall gravitational collapse to convert their gas into bouts of star formation. Within the solar circle, the mass of the Milky Way in GMCs is of order $\sim 10^9$\,M$_\odot$.  With a free-fall time of $4.35$\,Myr for gas at the typical average GMC density of 100\,cm$^{-3}$, this would yield a star formation rate of $230$\,M$_\odot$yr$^{-1}$; two orders of magnitude greater than the observed rate of  $\sim 1$--$2$\,M$_\odot$yr$^{-1}$ \citep{Chomiuk2001}. The GMCs are therefore either unbound structures and not dominated by their gravity, or they receive additional support from sources such as turbulence, magnetic fields, and/or stellar feedback.

Whether GMCs are gravitationally bound is difficult to determine. Both observations and global simulations that form clouds within a modelled galactic disc, suggest that GMCs are borderline bound, measuring a virial parameter $\alpha_{\rm vir} = 5\sigma_c^2 R_{\rm c}/(GM_c) \sim 1$, where $\sigma_c$, $R_c$ and $M_c$ correspond to the cloud velocity dispersion, mass and radius \citep{RomanDuval2010, Dobbs2011, Tasker2015, Jin2017}. Unfortunately, the errors in measuring $\alpha_{\rm vir}$ observationally are large, due to projection effects that force observers to integrate cloud spatial properties along the line-of-sight. Typical errors when measuring cloud properties are within a factor of two, but the uncertainty for the derived $\alpha_{\rm vir}$ value comprehensively straddles the range from unbound to bound \citep{Beaumont2013, Pan2015, Pan2016}. Additional uncertainties to $\alpha_{\rm vir}$ further arise due to the assumption that the clouds are spherical, of uniform density, and isolated structures \citep{Federrath2012}. Despite these problems, the average cloud size of $M_c \sim 10^5$\,M$_\odot$ and $R_c \sim 10$\,pc suggests that GMCs ought to be bound unless they have a velocity dispersion greater than that offered by thermal support. There must therefore be a source of turbulence affecting the evolution of GMCs and determining the rate of conversion from gas into stars.

The drivers for this turbulence may be internal or external (see \citet{Federrath2017} for a review of turbulence drivers). Internal drivers chiefly consist of thermal and kinetic feedback from the production of stars. This can both disrupt neighbouring collapsing regions of gas and also trigger collapse at the edge of expanding shells of hot gas. In \citet{Shima2017} we explored the impact of photoionising radiation in an idealised GMC and one whose structure was formed in a global galactic disc simulation. The increased heating of gas around photoionising stars increased the mass of the largest stars by suppressing fragmentation and freeing a greater amount of gas to be accreted. The combined pressure from the feedback could eventually disrupt the cloud and throttle star formation, but if the gas was extremely dense the feedback was unable to dominate over gravity. This dependence on the local environment was also found by \citet{Krumholz2010}, who argues that a high surface density effectively traps radiation which suppresses fragmentation to allow more massive stars to form. \citet{HarperClark2009} and \citet{Dale2014} also note the result of feedback is heavily dependent on cloud structure.

In contrast to such internal forces, the injection of turbulence to support GMCs may be drawn from the surrounding galactic environment. In global simulations without star formation, \citet{Jin2017} found a wide range of solenoidal and compressive turbulence modes were created within clouds due to their mutual interaction within the galactic disc. In similar simulations that compared runs with and without star formation and thermal feedback, \citet{Tasker2015} found that cloud properties were not strongly dependent on internal processes, supporting the idea that cloud evolution can also be driven by external conditions. This was further investigated by \citet{Fujimoto2014}, who found global structures such as spiral arms and a galactic bar could change the range in cloud properties by increasing the rate of cloud interactions. Such global-scale interference is supported by observations that see variations in GMCs and specific star formation rates within different galactic structures \citep{Koda2009, Meidt2013, Momose2010}

Whether internal or external drivers dominate GMC evolution may be reflected in the resulting star formation. If external processes are the key component to dictating a cloud's star formation history, then the stellar population may principally may be created in cloud collision events. \citet{Tan2000} and \citet{Fujimoto2014b} found that if cloud collisions triggered star formation, they could explain the empirical relation between gas surface density and the surface star formation rate (Kennicutt-Schmidt relation, \cite{Kennicutt1998}). Observations have also supported the notion that cloud interactions are common events. Super star clusters with masses $\sim 10^4$\,M$_\odot$ packed into a cluster of radius $\sim 1$\,pc have been observed to be associated with clouds colliding at velocities around $20$\,km\,s$^{-1}$ \citep{Furukawa2009, Ohama2010, Fukui2016}, as well as sites of high-mass star formation \citep{Torii2015}. Massive stars and clusters are difficult to form through isolated gravitational collapse, since the formation of massive stars should heat the local environment and prevent further accretion (acting as a cap on maximum star size) and disperse the surrounding gas. During a collision, gas is compressed at the interface between the two clouds, creating a high density shell of gas. The rise in density can potentially reduce the local Jeans mass to lead to a lower characteristic stellar mass \citep{Hennebelle2009}, but the collision can also promote the production of massive stars from an increase in the velocity dispersion or a faster formation timescale that allows longer accretion from the surrounding dense gas. If such collisions can be interpreted as compressive driving of turbulence, then we expect an enhanced star formation rate by factors ranging from a few to up to 10 \citep{Federrath2012, Federrath2016b}.

In simulations of colliding clouds between 7 - 15\,pc in size, \citet{Takahira2014} found that collisions produced massive star-forming cores and a play-off took place between the shock speed --which heightened core production-- and the duration of the shock moving within the cloud, which allowed cores to grow within the dense environment. Parsec-scale clouds in simulations by \citet{Balfour2015} and \citet{Balfour2017} agreed that the slower collisions were more efficient at forming massive stars in these smaller systems, while \citet{Wu2017} confirmed a star formation rate and efficiency higher by a factor of ten in colliding GMCs with added support from magnetic fields. Ultimately, these results agree that if cloud evolution is driven by collisional interactions and result in a tendency towards more massive stars, then the resulting stellar initial mass function (IMF) should be more top-heavy than if the cloud evolved in isolation.

In this paper, we explore the star formation from isolated and colliding GMCs and the evolution after energy injection by stellar feedback. Unlike previous simulations in \citet{Takahira2014} and also those considered by \citet{Wu2017}, we include a sink particle method into a grid based code that allows stars to gather gravitationally bound and collapsing gas from their surrounding environment. These criteria ensure that if the turbulence is high, a higher mass will be needed for the particle creation. Therefore, an increase of turbulence within the shock will lead to more massive stars. We also include output feedback energy from photoionisation, but we do not include magnetic fields. In Section~\ref{sec:method} we describe our method, Section~\ref{sec:results} shows our results for the isolated and colliding cases and the outcomes are summarized in Section~\ref{sec:conclusions}.

\section{NUMERICAL METHODS}
\label{sec:method}

Our simulations use the adaptive mesh refinement (AMR) hydrodynamics code, {\it Enzo}, with a three-dimensional implementation of the {\it Zeus} hydro-code to evolve the gas \citep{Enzo, Zeus}. The simulation box size is $120$\,pc along each side, covered by a root grid with $128^3$ cells, with two additional static meshes that are located at the position of the clouds throughout the simulation. An additional three levels of adaptive refinement are used, based on the requirement that the Jeans length must not fall below five cells. This limit is slightly larger than the four cells per Jeans length suggested by \citet{Truelove1997} as the minimum needed to prevent spurious numerical fragmentation. We note that in order to resolve turbulence on the Jeans scale, an even higher resolution of 30 cells per Jeans length may be required \citep{Federrath2011}. At the maximum refinement level, the cell size is 0.029\,pc. Gas at this level that continues to collapse risks violating the Jeans length resolution criteria. At this point, we introduce a sink particle algorithm as described in \citet{Federrath2010a} to convert gas into pressure-less particles that can continue to accrete gas and radiate (see below).

Non-equilibrium cooling down to 10\,K (the typical temperature of the dense gas in GMCs) is computed within {\it Enzo} by following nine atomic and molecular species,  ${\rm H, H^+, He, He^{+}, He^{++}, e^{-}, H^-, H_2}$ and ${\rm H_2^+}$ and supplemented with metal and molecular cooling using a data table from the CLOUDY photoionisation package \citep{Ferland1998, Smith2008}. CLOUDY assumes a solar hydrogen mass fraction and metallicity, but does presume that the gas will be optically thin; a weakness when considering gas within GMCs. The mean molecular weight is calculated from the species abundances, giving a value $\mu\sim$1.2. In addition to stellar feedback, the gas is heated by photoelectric heating which is proportional to the gas density as implemented in \citet{Tasker2011}. The gas is self-gravitating in all simulations, computed using a multi-grid solver to solve the Poisson equation.

\subsection{Star Formation and Feedback}

Stars are born in the simulation following a modified version of {\it Enzo}'s star cluster formation scheme described in \citet{popII}, using the sink particle criteria proposed by \citet{Federrath2010a} to treat gravitationally bound and collapsing gas objects. A sink particle is formed when the following criteria are met: (a) cell density is greater than a threshold value, $\rho_{\rm th}$, (b) the cell is on the finest level of refinement, (c) the gas within the control volume (i.e., the Jeans volume) is not within the accretion sphere of another sink particle, (d) the gas in the control volume is converging towards the center of the control volume (e) the local gravitational potential minimum is within the accretion sphere, (f) gas within the accretion sphere is Jeans unstable for collapse and (g) gas within the accretion radius is gravitationally bound.

The binding energy is calculated as the sum of the gravitational, kinetic and thermal energy. Our simulation does not include support from magnetic fields, so our initial sink mass is a lower limit. The radius of the accretion sphere and the threshold density are calculated from the Jeans length criteria as $r_{\rm acc} = \lambda_J/2 = 2.5\Delta x_{\rm min}$ and $\rho_{\rm th} = \pi c_s^2/(G \lambda_J^2)$, where $\lambda_J$ is the minimum Jeans length we can resolve with five cells at maximum refinement, $\Delta x_{\rm min}$ is the minimum cell size and $c_s$ is the sound speed. For gas at 10\,K, our simulation uses an accretion radius of $0.073$\,pc and a threshold density of $\rho_{\rm th} = 7.0\times 10^{-20}$\,g cm$^{-3}$, which is above the observed value ($10^4$\,cm$^{-3} \simeq 3\times 10^{-20}$\,g cm$^{-3}$) for star formation to occur \citep{Lada2010, Ginsburg2012, Kainulainen2014}.

When a sink is formed, gas over the threshold density is removed from the cell to create the initial particle's mass. The sink is given the average velocity of the neighbouring cells, avoiding the potential for a runaway particle. Once created, the sink can accrete gas over the threshold density that is within the accretion sphere and bound to that sink. The sink gathers mass for a local dynamical time (the free-fall time inside the accretion sphere) or until the sink mass hits 50\,M$_{\odot}$; a cluster mass that is liable to contain at least one massive star based on a Salpeter IMF between 1--100\,M$_\odot$. If another particle enters the accretion sphere during the accretion phase, the sink particles will merge if they are gravitationally bound. The resulting sink should therefore be thought of as a star cluster, not an individual star.

If the sink particle mass exceeds 20\,M$_{\odot}$ after the accretion has finished, the particle will emit ionising radiation. This is calculated using the adaptive ray tracing scheme implemented in {\it Enzo} that is described in \citet{Abel2002, Wise2011} and based on the HEALPix framework \citep{HEALPIX}. Each particle has an ionising luminosity of $10^{46.85} {\rm ph\,s^{-1}\,M_{\odot}^{-1}}$, which assumes solar metallicity and a Salpeter IMF between $1$--$100$\,M$_\odot$ \citep{Schaerer2003}. Notably within this range, the IMF gradient does not depend strongly on our model choice and the Chabrier IMF produces a similar distribution \citep{Chabrier2005}. The rays are assumed to be monochromatic with a mean ionising photon energy of $E_{\rm ph} = 20.84$\,eV. The gas is assumed to be predominantly hydrogen to give a heating rate of $E_{\rm ph} - E_{\rm H} (= 13.6\,{\rm eV})$. Since we ignore the radiative feedback from sink particles below 20\,M$_\odot$ for computational reasons, our feedback can be considered a lower limit.

\subsection{Initial Conditions}

\begin{figure}
 \begin{center}
   \includegraphics[width=16cm]{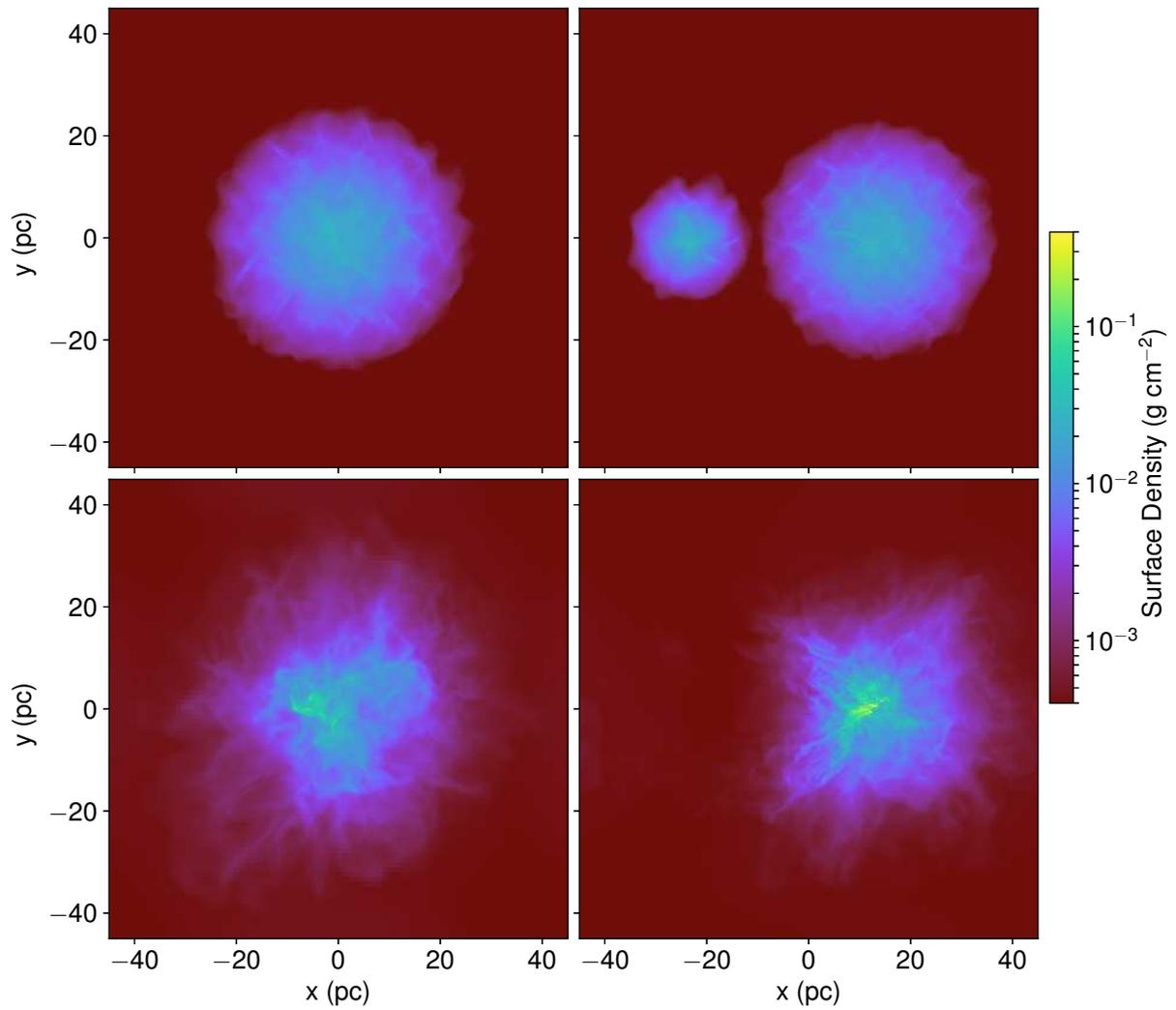}
 \end{center}
  \caption{(Top) Surface density of the isolated (left) and colliding clouds (right) after 0.5\,Myr and (bottom) at our main comparison time after 4\,Myr. Shown is the colliding case without feedback for a collision velocity of 10\,km s$^{-1}$.}
\label{fig:projection}
\end{figure}

We consider three sets of initial conditions. The first simulation is an isolated cloud, while the second and third sets are clouds with differing collision velocities. The mass of the isolated cloud is equal to the combined mass of the two colliding clouds.

The clouds are surrounded by low density ($1.4\times 10^{-24}$\,g cm$^{-3}$) background gas at $10^{4}$\,K. The clouds themselves have a Bonnor-Ebert density profile which takes the form of a hydrostatic, isothermal and self-gravitating sphere of gas that is confined by its external pressure \citep{Bonnor1956}. The origin of this profile is analytic, however observations have provided evidence of such profiles occurring in the Universe, such as the Bok Globule, B68 \citep{Alves2001}. The Bonner-Ebert sphere is unstable if its dimensionless radius, $\xi = r_{\rm c}/(c_{s} / \surd(4\pi G \rho_{0}))$ ($\rho_0$ is the central cloud density), exceeds a critical value of $\xi_{\rm crit} = 6.45$. Our spheres sit at the slightly higher value of $\xi = 7$ and therefore begin to collapse after the start of the simulation.

The cooling of the gas is initially compensated by an initial injection of turbulence that imposes a velocity field within the cloud. This is described in detail in \citet{Takahira2014, Shima2017}. In brief, a velocity power spectrum of $v_k \propto k^{-4}$ is added to the gas, corresponding to the expected spectrum given by Larson for GMCs \citep{Larson1981, MacLow1998} and appropriate for the supersonic turbulence seen in molecular clouds \citep{Federrath2013}. This creates a filamentary density structure \citep{Arzoumanian2011, Federrath2016a}, rather than a centralised collapse. The initial amplitude of the turbulence is dictated by the Mach number, $\cal{M} \equiv {\rm \sigma_c} / {\rm c_s}$ $= 1.0$, where $\sigma_c$ is the velocity dispersion inside the cloud and $c_s$ is the sound speed at the equilibrium temperature. Initially the Mach number is 1.0, but the cloud cools within a few Myrs, which is shorter than the crossing time ($\sim 10$\,Myr). After this, the turbulence is supersonic with a typical Mach number of 10.

\begin{table}
  \tbl{Idealized (Bonnor-Ebert) cloud initial parameters: radius, mass, temperature, average density, free-fall time and velocity dispersion.}{%
    \begin{tabular}{@{}lllll@{}}
      \hline
                      & isolated & small & large & \\
      \hline
      $r_c$           & 24.5   & 11    & 22    & pc \\
      $M_{c}$         & 5.5    & 1.1   & 4.4   & $\times 10^{4}$\,M$_{\odot}$ \\
      $T_{c}$         & 580    & 260   & 520   & K \\
      $\bar{\rho}$    & 6.1    & 13.6  & 6.8   & $\times 10^{-23}$\,g cm$^{-3}$ \\
      $t_{ff}$        & 8.5    & 5.7   & 8.0   & Myr \\
      $\sigma_{c}$    & 3.8 & 2.7 & 3.5 & km\,s$^{-1}$ \\
      \hline
    \end{tabular}}\label{table:cloud_value}
\end{table}

For the simulations of colliding clouds, we evolve the clouds in a stationary position for 0.5\,Myr to allow the development of the turbulent density structure. The smaller cloud is then given a bulk velocity of 10\,km s$^{-1}$ or 20\,km s$^{-1}$ in the direction of the larger cloud to form a head-on collision. These velocities were based on simulations performed by \citet{Takahira2014}, who found that slower velocities did not result in a strong shock front while a faster collision completed the interaction before the dense gas had time to collapse. This sweet spot of around $10 - 20$\,km s$^{-1}$ also appears to be supported by the observed collision velocities of clouds around super star clusters.

Figure~\ref{fig:projection} shows the surface density along the z-axis at 0.5\,Myr when the turbulent structure has been established (top panels) and at 4\,Myr (bottom panels) which is our main analysis time. The left-hand panels show the isolated cloud, while the right-hand panels show the colliding clouds with velocity 10\,km s$^{-1}$. Our analysis time for comparing the isolated and colliding clouds was chosen to be half the free-fall time of the isolated (and largest colliding) cloud. A summary of the cloud properties is shown in Table~\ref{table:cloud_value}.

\section{Results}
\label{sec:results}

To assess the impact of both the collision and feedback, this section will first look at the comparison of the isolated and colliding clouds where sink particles do not emit ionising radiation. The final part of this section will consider the addition of this feedback.

\subsection{Gas Structure}

\begin{figure}
 \begin{center}
   \includegraphics[width=8cm]{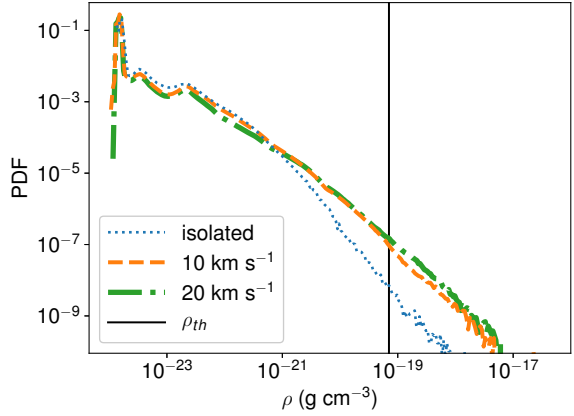}
 \end{center}
  \caption{The probability distribution function for the isolated (blue-dot) and two colliding cloud systems (orange-dash and green-dot-dash) at 4\,Myr. The clouds involved in the collision have wider PDFs due to the rise in compressive turbulence.}
\label{fig:pdf}
\end{figure}

The range of densities within the cloud gas can be seen in the one-dimensional density probability distribution functions (PDF) in Figure~\ref{fig:pdf}, for our analysis time of 4\,Myr. The blue dotted line shows the isolated cloud, while the orange dashed and green dot-dashed lines are for the clouds that collided with 10\,km s$^{-1}$ and 20\,km s$^{-1}$, respectively. The black vertical line shows the density threshold for the sink particle creation. Gas to the right of this line is therefore eligible to be converted into stars, assuming it is gravitationally bound to a sink (as described in section~\ref{sec:method}).

At the sink threshold, the colliding cloud systems both have over a factor of ten more dense gas than in the isolated case. This extends into a power-law tail that is expected when gravity starts to dominate over turbulence \citep{Klessen2000, Kritsuk2011, Federrath_Klessen, Schneider2013, Kainulainen2014}. The excess dense gas is from the interface of the collision which creates a bow-shaped shock as the smaller cloud penetrates the body of the larger GMC (see \cite{Takahira2014} for images of the evolution of the collision). The result is a substantially broader PDF profile, demonstrating that compressive turbulence is being driven by the collision of the clouds \citep{Federrath2008, Federrath2010b}. This is expected to lead to an enhanced star formation rate \citep{Federrath2012}. Notably, gas with densities higher than the sink formation threshold is not being instantly converted into sink particles because of the sink creation criteria of bond and collapsing gas introduced by \citet{Federrath2010a} (see method section). This suggests that while dense, the turbulence is kinetically supporting the gas against collapse and preventing it being immediately available for sink creation. Larger cores with masses able to dominate the turbulence should therefore be favoured and will accrete from the surrounding gas to form the massive clusters seen in observations of cloud collisional systems \citep{Fukui2016}. This will be shown explicitly below. Between the two collision velocities, the gas distributions are  similar, with the faster impact producing slightly more dense gas beyond the sink threshold.

\subsection{Star Formation}

\begin{figure}
 \begin{center}
  \includegraphics[width=8cm]{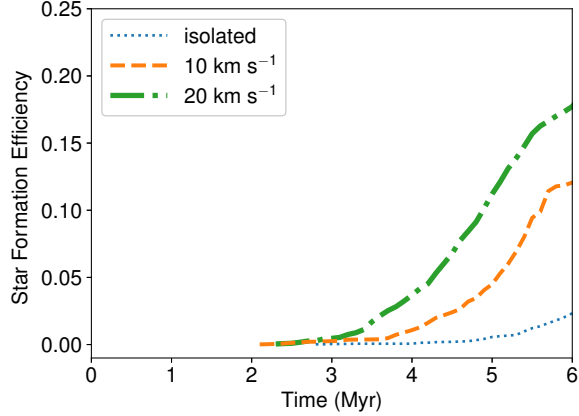}
 \end{center}
\caption{The star formation efficiency for the simulations without feedback. The blue dotted line shows the isolated cloud case, the orange dashed line is for the clouds colliding at 10\,km s$^{-1}$ and the green dot-dashed line is for the faster 20\,km s$^{-1}$.}
\label{fig:SFE}
\end{figure}

How actively the cloud converts its gas into stars can be measured by looking at the cloud's star formation efficiency (SFE), defined as $\epsilon(t) = M_{\rm star}(t) / M_{\rm cloud} (t = 0)$, where $M_{\rm star}$ is the total mass in stars at time $t$, and $M_{\rm cloud}$ is the total cloud mass (for the colliding case, for both clouds) at the start of the simulation. The evolution of the SFE is plotted in Figure~\ref{fig:SFE} for the isolated (blue-dot) and two colliding systems (orange-dash and green-dot-dash) without photoionising feedback. 

Initially, the gas is not dense enough to form sink particles. Thermal support decreases as the gas cools, then stars start to form earlier than the global collapse timescale ($t_{ff}\sim$8\,Myr) due to local compression by turbulence. The first star forms in the isolated cloud at 2.7\,Myr, but very few stars are formed until after 5\,Myr where the cloud's gravity finally overwhelms the decaying turbulence and begins to collapse. This leads to the global collapse of the isolated cloud. Without a fresh form of support from stellar feedback, the number of stars in the isolated cloud increases rapidly, reaching an efficiency of about 2\% by 6\,Myr.

The colliding clouds begin star formation earlier, starting at 2.0\,Myr and 2.2\,Myr for the 10\,km\,s$^{-1}$ and 20\,km\,s$^{-1}$ collision velocities respectively. The clouds begin their collision at 1\,Myr for the 10\,km s$^{-1}$ and 0.8\,Myr for the 20\,km s$^{-1}$ cases, so star formation begins about 1\,Myr after the shocked interface begins to form. Sink particles form rapidly in the dense shock front, producing a significantly higher star formation rate (the gradient of the SFE curve) than for the isolated case. The collision at 20\,km s$^{-1}$ produces a denser shock front than the slower collision, increasing the stellar production rate still further. At 4\,Myr, the 10\,km s$^{-1}$ collision case is forming stars 18 times faster than for the isolated cloud, while the 20\,km s$^{-1}$ collision case is 53 times as rapid. This demonstrates how sensitive the star formation rate is to gas compression \citep{Federrath2012, Federrath2016b}. By 6\,Myr, the colliding clouds have a SFE of between 10 - 20\%; a factor of 10 higher than the isolated cloud.

Observations of GMCs suggest that the SFE should be of order a few percent (for example, see Table 3 in \cite{Federrath_Klessen}, which lists the SFE for various clouds in the Milky Way.) This would initially appear to agree better with the isolated cloud than the collision cases. However, true GMCs are typically more extended structures than our idealised Bonnor-Ebert spheres. Real clouds are part of the Galaxy's dynamic ISM and thus subject to turbulent perturbations and tidal interactions that create extended, elongated structures. A cloud collision is therefore likely to involve only part of the cloud. Examples of this can be seen in \citet{Fukui2017}, who shows the observational maps of cloud collisions within the LMC or in the numerical simulations of \citet{Jin2017}. Collisions are therefore likely to affect a smaller fraction of the cloud's volume than in this idealised case, driving down the total SFE when averaged with more quiescent regions.

\begin{figure}
 \begin{center}
  \includegraphics[width=16cm]{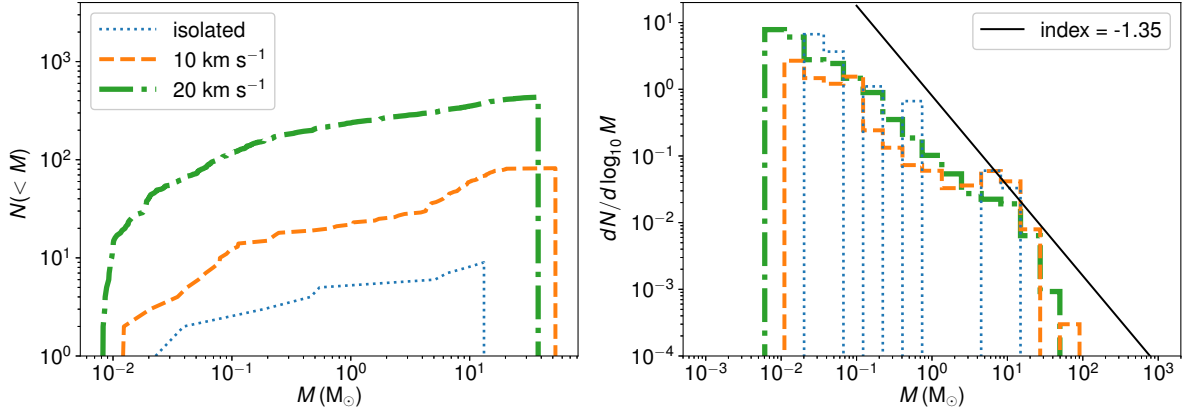}
 \end{center}
  \caption{The cumulative stellar mass distribution (left) and stellar mass function (right) for the isolated and two colliding cloud cases without photoionising feedback at 4\,Myr. Black solid line shows the gradient of the Salpeter IMF.}\label{fig:mass_function}
\end{figure}

The resulting star formation can be seen in the cumulative mass in star particles shown in the left-hand panel of Figure~\ref{fig:mass_function} for the simulations without feedback by 4\,Myr. The collisional simulations form more low mass stars and massive star clusters than the isolated case, confirming that turbulence in the collisional shock front also acts as kinetic support to favour the production of massive stars. The largest cluster formed in the isolated cloud is 13\,M$_{\odot}$, while the collision at 10\,km s$^{-1}$ yields a maximum cluster mass of 53\,M$_{\odot}$ and a slightly lower 38\,M$_{\odot}$ for the 20\,km s$^{-1}$. While the maximum mass is reduced in the faster collision compared to the slower interaction, the total stellar production is greater, with the 20\,km s$^{-1}$ creating around five times more sink particles than in the 10\, km s$^{-1}$ collision. The smaller size of the maximum mass star cluster is likely due to the faster shock front providing less time to accrete gas, a factor noted by \citet{Takahira2014} as the reason why steadily faster shocks do not form ever larger clumps. By contrast, the isolated cloud case created only 9 sink particles during the first 4\,Myr, compared with 83 and 434 in the colliding runs.

The distribution of the stellar masses of the sink particles can be seen in the mass function in the right-hand panel of Figure~\ref{fig:mass_function} for the same runs. The black solid line in this figure shows a power-law relation with an index of $-1.35$, which is the prediction of the Salpeter IMF, described by $dN/d\log_{10}{M} \propto m^{-\chi}$ for $\chi = 1.35$. Since our simulations really only resolve star clusters, the mass function we plot is best interpreted as a molecular core mass function and not as the stellar IMF. Despite significant differences in the total number of stars and the maximum cluster mass, both the isolated and colliding cloud runs show a similar gradient that is consistent with the slope of the Salpeter IMF. It is worth noting that the isolated cloud has only nine sink particles, so the distribution for this run is relatively scarce. However, the particles formed do appear to follow the same trend as for the colliding cases.

At the high-mass end, the colliding clouds both show extensions beyond the isolated case, indicating the effectiveness of forming high-mass stars during a collision. There is evidence in this region that the IMF may steepen. This fall-off likely corresponds to star formation that occurs only in the densest part of the shock interface.

\subsection{The Effect of Stellar Feedback}

While the cloud collision provides an external source of stirring to the GMC gas, the stars themselves will add energy to their surrounding medium. To explore this, we include photoionising feedback as described in Section~\ref{sec:method} in the isolated cloud simulation and the collisional case at 10\,km s$^{-1}$ impact velocity.

\begin{figure}
 \begin{center}
  \includegraphics[width=8cm]{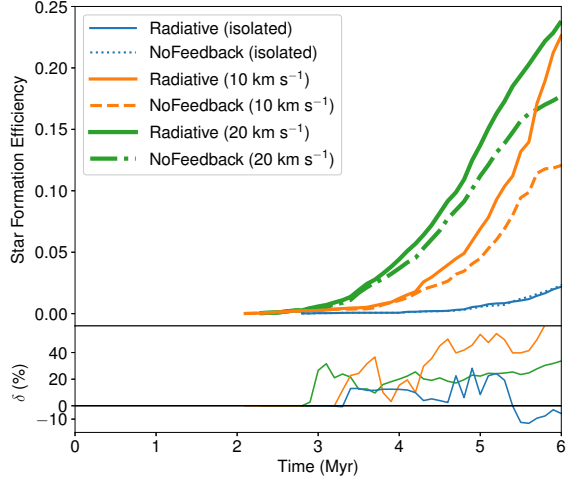}
 \end{center}
  \caption{Star formation efficiency for the isolated (blue / blue dot-dot), 10\,km s$^{-1}$ collision (orange / orange-dash) and 20\,km s$^{-1}$ collision simulation (green / green-dot-dash) when photoionising feedback is included. Feedback runs are denoted by the solid lines. The ratio of the difference between non-solid (non-feedback) and solid (feedback) lines are also shown in the lower panel, with a +/- value indicating the positive/negative effect of feedback.}\label{fig:SFE_feedback}
\end{figure}

The evolution of the star formation efficiency when feedback is included is shown in Figure~\ref{fig:SFE_feedback}. The dashed and dotted lines show the same result for the non-feedback case as in Figure~\ref{fig:SFE}, with blue-dots showing the result for the isolated cloud and orange-dash and green-dot-dash for the 10\,km s$^{-1}$ and 20\,km s$^{-1}$ collisional case, respectively. The solid blue line shows the isolated cloud SFE when photoionising feedback is included, while the solid orange and green lines show the feedback results for the colliding clouds. The bottom panel of the figure shows the difference between the equivalent non-feedback and feedback runs, where $\delta$ is defined as $\delta (t) = {\rm(SFE_{\rm Feedback}(t) - SFE_{\rm No Feedback}(t))/{SFE_{\rm No Feedback}(t)} \times 100}$\%. A positive value of $\delta$ corresponds to feedback promoting the star formation in the cloud, while a negative value means that star formation is being suppressed.

In both the isolated and colliding cases, the effect of feedback is generally to promote star formation over the 6\,Myr. This effect was previously seen in \citet{Shima2017} in cases where the feedback had a chance to act before the gas became so dense that the self-gravity dominated over the pressure from the feedback. There, the radiation heated the gas to suppress fragmentation and allowed the gas to be accreted to form more massive stars. This persisted until the cloud was dispersed by the feedback, leading to a subsequent drop in star formation.

In the isolated cloud, we repeat this effect. Feedback initially makes a positive difference and raises the SFE. As the dense clumps within the cloud begin to be dispersed around 5.5\,Myr, the star formation drops compared to the non-feedback case were collapse continues unabated.

When feedback acts during the cloud collision, the effect is uniformly positive, with feedback aiding stellar production. This is from the photoionisation heating of the dense gas within the shocked collisional interface, preventing fragmentation and allowing still larger stars to form. This agrees qualitatively with the smaller-scale simulations of ionising feedback by \citet{Peters2010, Peters2012}, who simulated collapsing molecular cores with radiative feedback and found that the heated gas boosted the Jeans mass to form massive star clusters. While the feedback there could be both positive and negative, it was primarily positive when averaged over longer timescales. This boost in stellar production is stronger in the 10\,km\,s$^{-1}$ case than in the faster 20\,km\,s$^{-1}$ collision. As we will see below, this is due to the speed of the shock front compared to that of the  expanding HII region.

\begin{figure}
 \begin{center}
  \includegraphics[width=14cm]{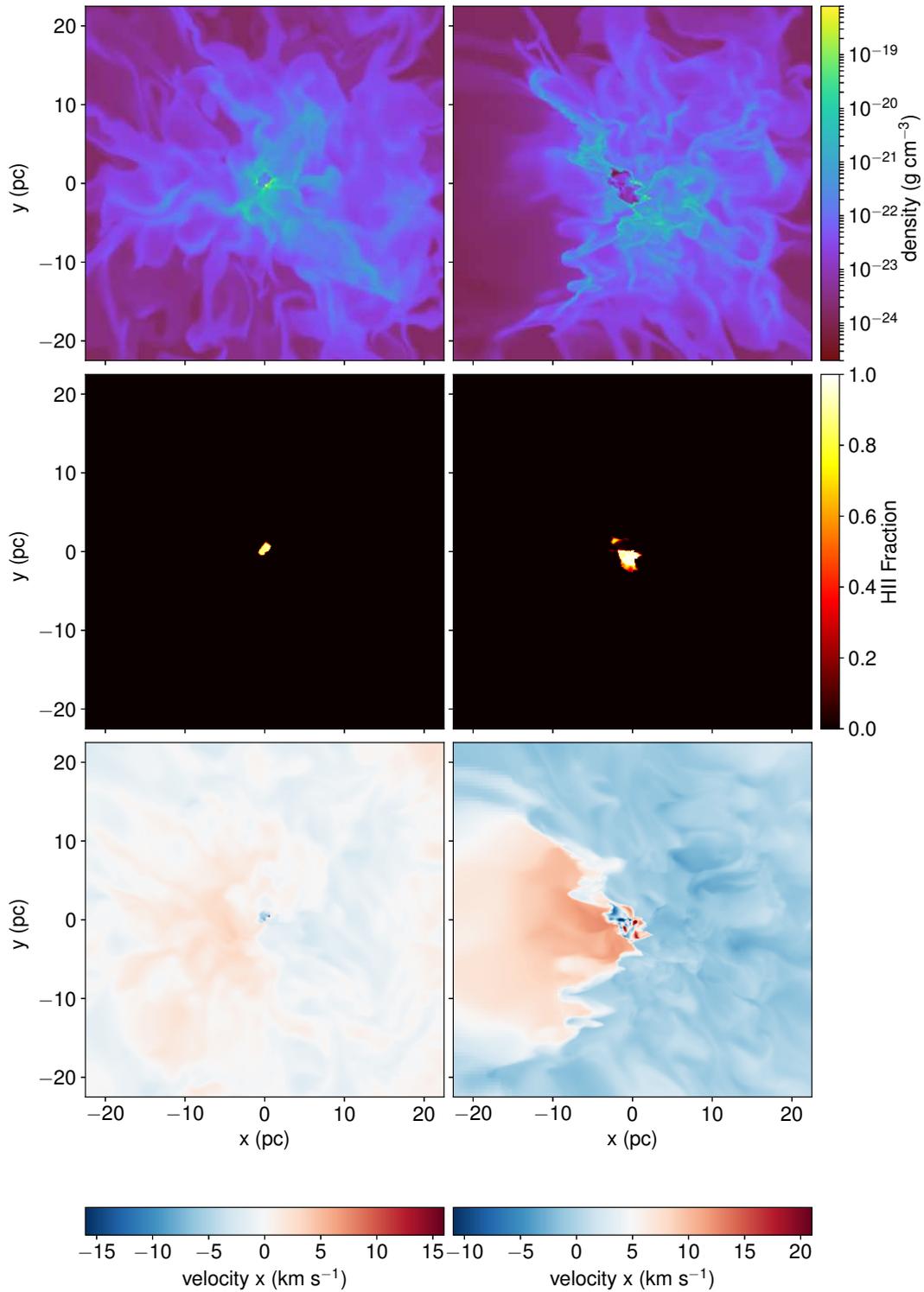}
 \end{center}
  \caption{The gas density (top), HII fraction (middle), and x-velocity (bottom) on a slice for the feedback runs in the isolated cloud case (left) and 10\,km s$^{-1}$ collisional case (right). Images are centered around the most massive sink particle a few Myr after it begins to emit radiation. The isolated cloud is shown at 5.2\,Myr, while the colliding case is at 3.7\,Myr as the sink formation time differs between these runs. The off-set in the velocity colourbar matches the shock propagation speed in the colliding cloud case of 5\,km s$^{-1}$.}\label{fig:panel}
\end{figure}

\begin{figure}
 \begin{center}
  \includegraphics[width=8cm]{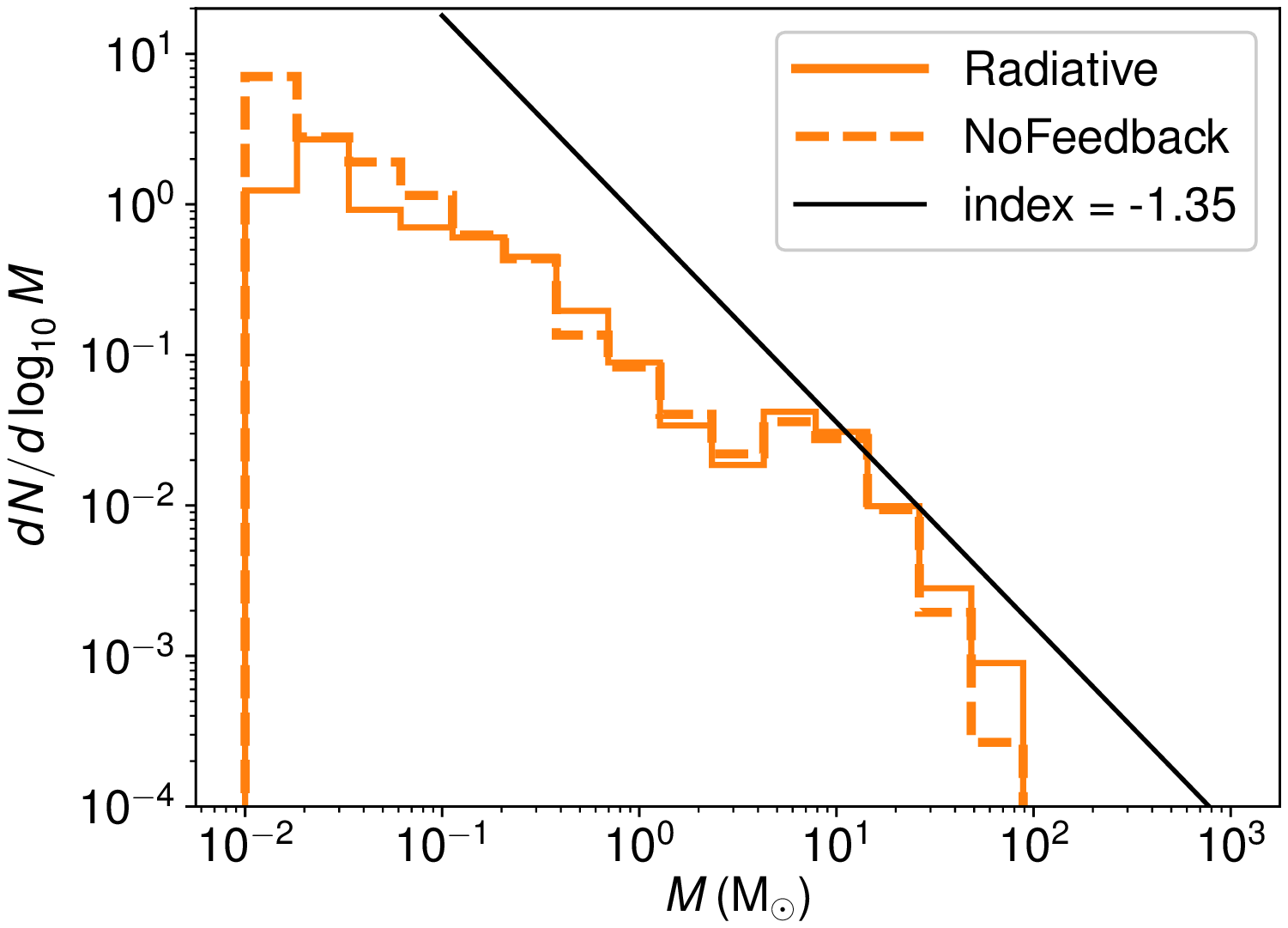}
 \end{center}
  \caption{Stellar mass function at 4.5\,Myr for the 10\,km s$^{-1}$ colliding cloud case with and without photoionising feedback.}\label{fig:massfunc_feedback}
\end{figure}

Exactly why the feedback can have different effects on the cloud SFE can be seen visually in a close-up image of the gas around the most massive sink particle a few Myr after it has begun to emit radiation in Figure~\ref{fig:panel}. The left-hand column of images shows the gas around the massive sink in the isolated cloud case, while the right-hand images shown the collisional case for a velocity of 10\,km s$^{-1}$. In the middle row of images showing the HII fraction (HII mass divided by hydrogen mass), a bubble of hot ionised gas can be seen expanding around the photoionising sink. The total HII mass is 0.003\,M$_\odot$ in the isolated cloud case and 6.2\,M$_\odot$ in the collisional case. The top row shows the gas density in a slice at the sink's position, while the bottom row shows the velocity in the direction of the collision. In the collisional case, the velocity axis has been shifted to match the shock propagation speed. As the formation of the sink particles differ between simulations, the isolated case shows the most massive sink forming at 5.2\,Myr, while the collisional case is at 3.7\,Myr.

The isolated cloud forms its massive stars close to the cloud center, as gas is collapsing towards this region as the initial turbulent support decays. As the massive star cluster forms and begins to photoionise, the radiation heats the surrounding layers of gas and prevents further fragmentation in this region. This suppresses the formation of smaller stars, but allows the gas to be accreted onto the massive star cluster. The mass of the star clusters forming in the simulation therefore increases, as in \citet{Shima2017}. Eventually, the heat from the centrally concentrated radiating stars counters the gravitational collapse and further accretion and the star formation begins to slow. This results in the negative effect we see in Figure~\ref{fig:SFE_feedback} around 5.5\,Myr.

In the case of the colliding clouds, the gas structure is markedly different. The first massive sinks form in the shock generated at the collision interface. As the shock continues to move forward, the HII region expands behind and in front of the shock front. The shock propagation speed for the collision at 10\,km s$^{-1}$ is approximately 5\,km s$^{-1}$ through the cloud, but with a thermal temperature of $\sim$ 10$^{4}$\,K, the maximum expanding velocity due to the pressure gradient is around 10\,km s$^{-1}$ corresponding to the sound speed within the HII region. This allows the HII region to stay within the dense gas piling onto the shock front, continuing to suppress fragmentation in this very dense region and allowing the gas to be accreted onto the massive sinks. Since the high density of the shock front is harder to dispel than the gas in the isolated case, feedback continues to have a positive impact on the SFE. This can be seen quantitatively in the mass function for the run in Figure~\ref{fig:massfunc_feedback}. The orange solid line corresponds to the simulation with radiation feedback, while the dashed line is for the non-feedback case for the cloud collision at 10\,km s$^{-1}$, both taken at 4.5\,Myr. The photoionisation results in more high-mass star particles being created as fragmentation into smaller stars is suppressed. While the trend for the stellar mass function for the faster cloud collision at 20\,km\,s$^{-1}$ is the same, the overall effect is more modest. In the bottom panel showing the net effect of feedback on Figure~\ref{fig:SFE_feedback}, the green line is below the orange line. This is again due to the swiftness of the shock front passing through the cloud. With the higher propagation speed, the expanding HII region is outstripped and can have less effect on the star-forming gas. As we have seen, the reduced time of the shock front within the cloud also lowers the maximum mass of the sinks, additionally lowering the impact of the feedback.

Previous simulations have suggested that expanding HII regions could still promote the formation of smaller stars in a `collect and collapse' scheme, where stars form around the edge of the expanding bubble. However, as in \citet{Shima2017}, we find no evidence of that mechanism in play within our simulations, neither within the isolated nor the collisional case.

\section{Conclusions}
\label{sec:conclusions}

We compared stars forming in idealised isolated and colliding cloud cases at two different velocities with a sink particle method for forming stellar clusters and also emitting photoionising radiation.

The effect of the cloud collision was to increase the total stellar mass formed and also the maximum size of the star cluster born in the simulation, compared with the isolated cloud case. The collision created a dense shock front and compressed the gas within the colliding clouds.  The added density provided a larger reservoir of gas for star formation, while the compressive turbulence promoted the production of more high mass stars. The higher collision velocity increased the total stellar mass due to the production of a denser shock front, but the slower velocity speed produced the highest individual cluster mass and strongest response to feedback, due to the reduced speed increasing the time for the sink particle to accrete.

The overall gradient of the stellar mass distribution was not strongly different between the isolated and colliding cases and remains consistent with a Salpeter IMF for star clusters up to 10\,M$_\odot$. At the higher mass clusters achieved during the cloud collision simulations, the power-law relation appears to steepen. This may suggest that the biggest stellar clusters can only be formed during a collision between clouds and not through regular gravitational collapse.

The addition of photoionising feedback affected the isolated and colliding clouds differently, due to differences in the structure of the gas. In both cases, the radiation initially suppressed fragmentation, throttling the production of smaller stars but allowing more gas to be accreted onto the larger star clusters. This created a positive impact on the SFE. In the isolated case, radiation from the centrally concentrated star formation began to counter the gravitational collapse after several Myr. This ultimately slowed the production of stars and throttled the cloud SFE. In the collisional case, the expanding HII region from the massive star clusters was able to keep pace with the shock front as it travelled through the cloud. The particularly dense gas inside the shock front was not dispersed by the radiation, which continued to have a positive impact on the SFE.

The overall effect of the collision is therefore to boost the production of high-mass stars. Providing the shock speed is comparable to the sound speed within the resulting HII regions, feedback can be a positive force on star formation triggered in collisions, while it ultimately curtails star formation in the less dense isolated system.

\begin{ack}
Numerical computations were carried out on Cray XC30 at the Center for Computational Astrophysics (CfCA) of the National Astronomical Observatory of Japan. The authors would like to thank the YT development team for many helpful analysis support \citep{Turk2011}. EJT was partially supported by JSPS Grant-in-Aid for Scientific Research Number 15K0514. C.F. acknowledges funding provided by the Australian Research Council's Discovery Projects (grants DP130102078 and DP150104329). C.F. thanks the Juelich Supercomputing Centre (grant hhd20), the Leibniz Rechenzentrum and the Gauss Centre for Supercomputing (grants pr32lo, pr48pi and GCS Large-scale project 10391), the Partnership for Advanced Computing in Europe (PRACE grant pr89mu), and the Australian National Computational Infrastructure (grant ek9), as well as the Pawsey Supercomputing Centre with funding from the Australian Government and the Government of Western Australia.
\end{ack}


\end{document}